\documentclass[RNAAS]{aastex63}

\shorttitle{GMRT observations of  Tau Boo and 55 Cnc}
\shortauthors{Narang et al.}

\begin{document}

\title{GMRT observations of the exoplanetary systems $\tau$ Bo\"otis and 55 Cancri}

\correspondingauthor{Mayank Narang}
\email{mayank.narang@tifr.res.in}

\author[0000-0002-0554-1151]{Mayank Narang}
\author[0000-0002-3530-304X]{Manoj, P. }
\affiliation{Department of Astronomy and Astrophysics Tata Institute of Fundamental Research \\
Homi Bhabha Road, Colaba, Mumbai 400005, India}
\author[0000-0001-5356-1221]{C. H. Ishwara Chandra}
\affiliation{National Centre for Radio Astrophysics, Tata Institute of Fundamental Research \\ Pune University Campus, Pune, 411007, India}

\begin{abstract}
We present archival Giant Metrewave Radio Telescope (GMRT) observations of two exoplanetary systems, $\tau$ Bo\"otis, and 55 Cancri, at 610 MHz and 150 MHz, respectively. Theoretical models predict these systems to have some of the highest expected flux densities at radio wavelengths.  Both $\tau$ Bo\"otis and 55 Cancri have been previously observed at low frequency ($\sim$ 30 MHz) with Low-Frequency Array (LOFAR) \citep{Turner20}. $\tau$ Bo\"otis shows tentative signatures of circularly polarized emission at 30 MHz, while no emission was detected from 55 Cancri. We do not detect radio emission from both the systems, but the GMRT observations set $3\sigma$ upper limits of 0.6 mJy at 610 MHz for $\tau$ Bo\"otis and 4.6 mJy at 150 MHz for 55 Cancri. The sensitivity achieved at 610 MHz in these observations is comparable to some of the deepest images of an exoplanet field.

\end{abstract}
\keywords{Stars: planetary systems -- Stars: individual: $\tau$ Bo\"otis -- Stars: individual: 55 Cancri -- Techniques: interferometric radio }

\section{Introduction} \label{sec:intro}
The magnetic field of a close-in planet (orbital period $\leq$ 10 days) can interact with either the magnetic field of the host star or the stellar wind leading to star-planet-interaction (SPI).  This can result either in an enhancement of UV emission from the star or can lead to radio emission either from the star or planet  \citep[e.g.,][] {Griess17, Sh20, V20, Narang21}.   Recently \cite{Turner20} have reported tentative detection of highly circularly polarized radio emission from two exoplanetary systems, $\tau$ Bo\"otis and $\upsilon$ Andromedae, using the Low-Frequency Array (LOFAR) telescope at low frequencies ($\sim$ 30 MHz). They also observed a third system, 55 Cancri; but, no radio emission was detected from it. 

Here, we report our analysis of archival Giant Metrewave Radio Telescope (GMRT) observations of the systems $\tau$ Bo\"otis and 55 Cancri to search for evidence for SPI in the radio domain.  $\tau$ Bo\"otis (hereafter $\tau$ Boo), is an F7 V main-sequence star hosting a $5.9_{-0.20}^{+0.35}$~$M_J$ mass planet, $\tau$~Boo~b, with an orbital period of 3.3 days \citep{Lockwood14}.  55 Cancri~A (hereafter 55 Cnc), on the other hand, is an M4.5 star hosting five planets, with the closest planet 55 Cnc~e having a period of only 0.73 days and a mass of 8.63~$\pm$~0.35~$M_\oplus$ \citep{Winn11}. These two systems are thought to be among the most promising systems to detect radio emission from due to star-planet-interaction \citep{Griess17}.

\section{Observations and data reduction}
The GMRT observations of $\tau$ Boo at 610 MHz were carried out on 2009 July 21  and 2009 August 22 . On  2009 July 21, $\tau$ Boo was observed for 34 mins. 3C286 was used as the gain calibrator and was observed twice, once at the beginning (for 10 mins) and once at the end of the observation (for 13 mins).  The phase calibrator used was the Seyfert 2 Galaxy 1347+122, which was observed in a loop with the science target with  5 mins on the science target and 4 mins on the phase calibrator. For the second observation, on 2009 August 22, $\tau$ Boo was observed for 39 mins. 3C147   was used as the gain calibrator at the beginning of the observation (observed for 15 mins), and 3C286 was used as the gain calibrator at the end of the observations (observed for 16 mins). The phase calibrator used was 1347+122 and was observed in a loop with the science target.        

The observations of 55 Cnc (at 150 MHz) were carried out on 2005 August 22  and 2013 February 27. On 2005 August 22, 55 Cnc was observed for 1.5 hrs with 3C147 used as both the gain and phase calibrator. The system 55 Cnc was also observed in 2013 for 6 hrs. The gain calibrators 3C147 and 3C286 were observed at the beginning (3C147 for 21 mins) and at the end of the observations (3C286 for 21 mins). 0834+555 was used as the phase and bandpass calibrator and was observed in a loop for 6 mins on 0834+555 and 39 mins on the science target.  

All observations were reduced using the Source Peeling and Atmospheric Modeling (SPAM) pipeline \citep{Intema14}. Further analysis of the images was carried out with  Common Astronomy Software Applications (CASA) package.

\section{Results and Discussion}

In Figure 1 (a), we show the GMRT observations of $\tau$ Boo on 2009 July 21, and in Figure 1 (b), the observations from 2009 August 22 are shown.   In Figure 1 (c), the 150 MHz GMRT observation of the 55 Cnc from 2005 August 22  is shown, while Figure 1 (d) shows the observation of 55 Cnc from 2013 February 27 . Figure~1~(a) and Figure~1~(b), show a bright radio source $\sim$~4~arcmins away from $\tau$ Boo, with an integrated flux density of 111 $\pm$ 2 mJy at 610 MHz.

Though we have some of the deepest images of an exoplanet field, no radio emission was detected. For $\tau$ Boo, at 610 MHz, we achieved a $3\sigma$ upper limit of 0.8 mJy on 2009 July 21  and an $3\sigma$ upper limit of 0.6 mJy on 2009 August 22. For the system 55 Cnc at 150 MHz, we achieved a $3\sigma$ upper limit of 20 mJy on 2005 August 22 and $3 \sigma$ upper limit of 4.6 mJy on 2013 February 27 (a factor of 6 deeper than previous observations at 150 MHz). At 610 MHz, our $3\sigma$ upper limit values are comparable with the deepest observations of exoplanetary systems at that frequency range \citep{Griess17, Narang21}. 

There are several possible reasons for the null detection. The emission from the system could be beamed and only observable at a particular phase of the planet's orbit \citep[see][]{Narang21}. It could also be that the observations were made at a frequency away from the frequency of peak emission. For planets in the 55 Cnc system, the estimated cyclotron frequency is  $\leq 100$ MHz \citep{Griess17}. However, recent discoveries have pointed out that in the case of M dwarfs such as 55 Cnc, the emission frequency depends on the magnetic field of the host star and not the planet \citep{Prez20, Vedantham20}. In such a scenario, the emission from 55 Cnc would be close to 150 MHz. However, the system 55 Cnc was only observed for 6 hrs continuously, which is $\sim$ only one-third of the orbital period of 55 Cnc~e, the innermost planet. Deeper observations for a longer period are necessary to rule out emissions from 55 Cnc.  

\section{Summary}
We analyzed the archival GMRT observations of two exoplanetary systems, $\tau$ Boo and 55 Cnc. We do not detect any radio emission from either of the systems. At 150 MHz for 55 Cnc, we place a $3\sigma$ upper limit of 4.6 mJy on the quiescent component of the radio emission. For $\tau$ Boo at 610 MHz, we achieve a $3\sigma$ upper limit of 0.6 mJy on the quiescent component of the radio emission.

\begin{figure*}

\includegraphics[width=0.5\linewidth]{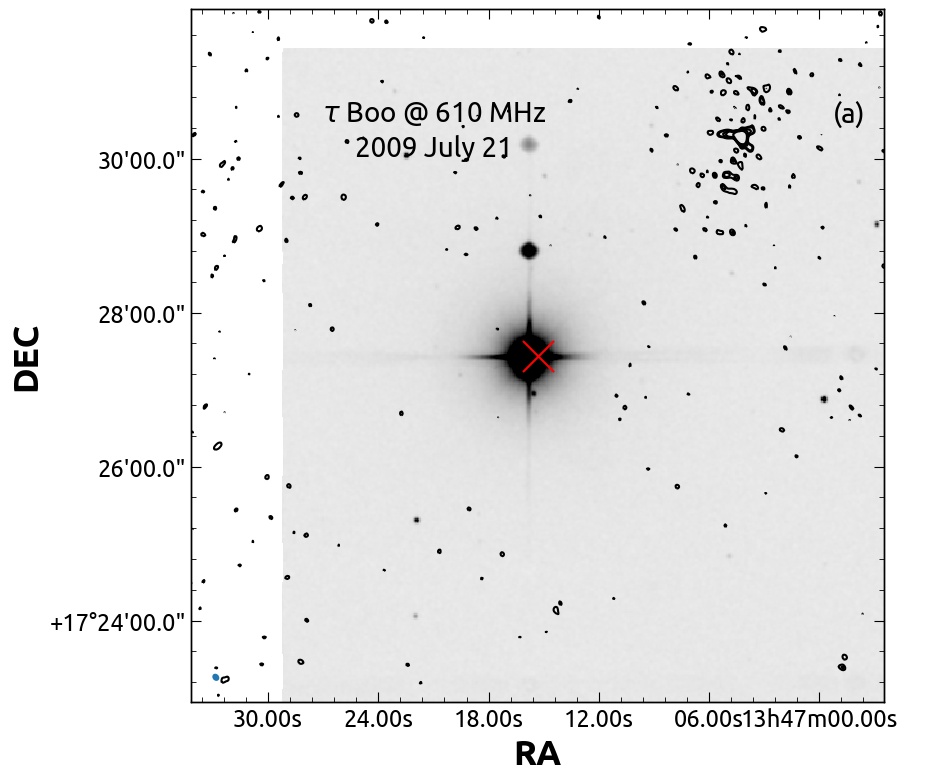} \includegraphics[width=0.5\linewidth]{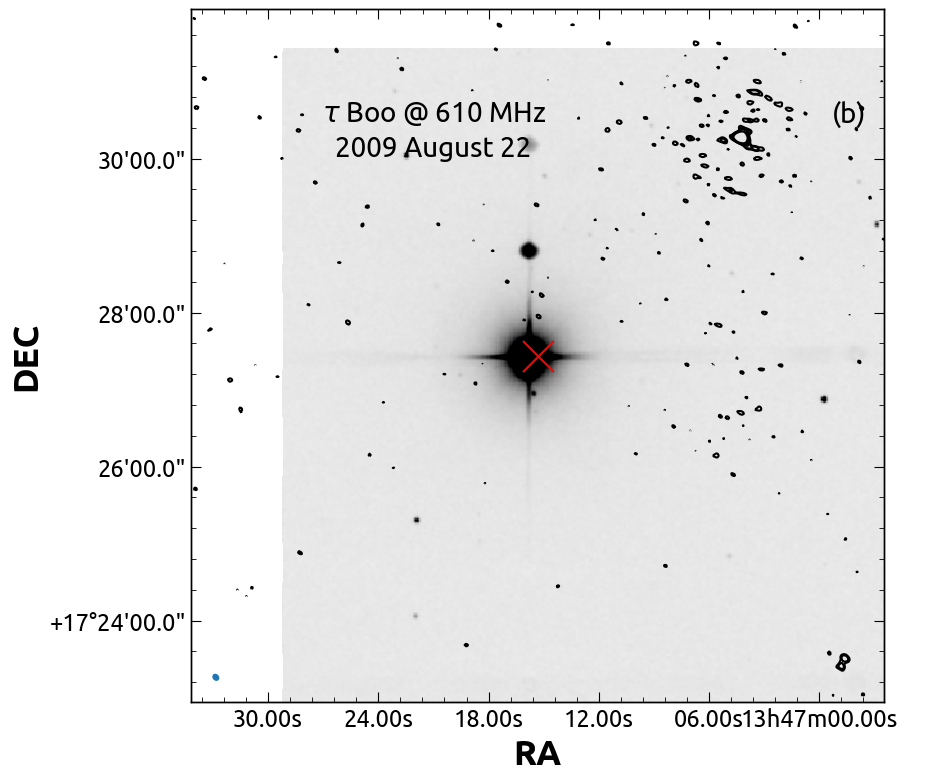}
\includegraphics[width=0.5\linewidth]{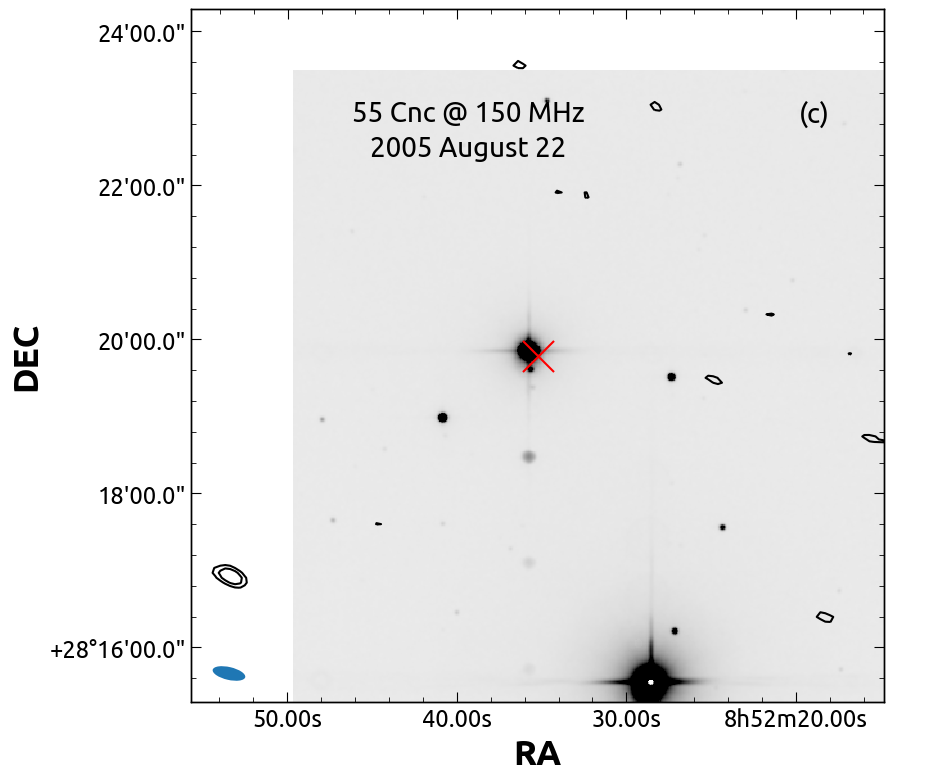} \includegraphics[width=0.5\linewidth]{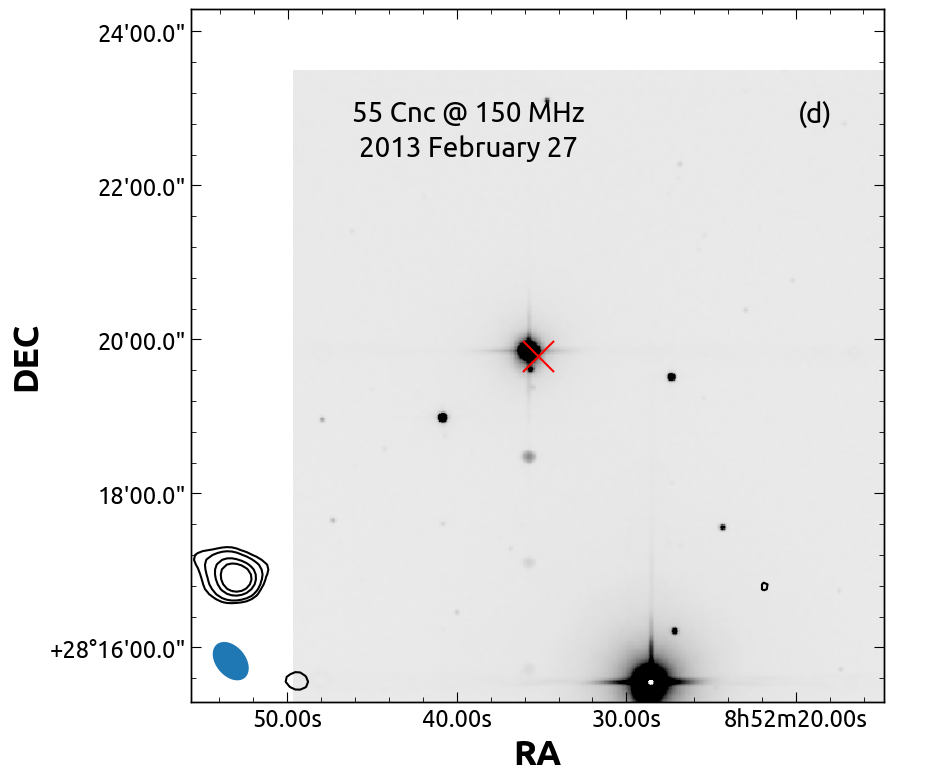}
\caption{ The GMRT image of the $\tau$ Boo field at 610 MHz on (a) 2009 July 21 and (b) 2009 August 22 and of 55 Cnc at 150 MHz  on (c) 2005 August 22  and (d) 2013 February 27  overlaid on the 2MASS J band image in greyscale. The red cross marks the $Gaia$ EDR3 \citep{Gaiae3} position of the host star. The contours plotted are 5, 10, 15, and 30~$\times\;\sigma$. The beam is shown as a blue ellipse at the bottom left corner. }
\label{fig1}
\end{figure*}

\section{Acknowledgement}
This work is based on observations made with the Giant Metrewave Radio Telescope, which is operated by the National Centre for Radio Astrophysics of the Tata Institute of Fundamental Research and is located at Khodad, Maharashtra, India. We thank the GMRT staff for efficient support to these observations. We acknowledge the support of the Department of Atomic Energy, Government of India, under the project  We acknowledge support of the Department of Atomic Energy, Government of India, under Project Identification No. RTI4002.

\bibliographystyle{aasjournal}

\end{document}